# Genomic prediction: progress and perspectives for rice improvement

Running title: Genomic prediction for rice breeding

Jérôme Bartholomé[1,2,3]*, Parthiban Thathapalli Prakash[3], & Joshua N. Cobb[4]

[1] CIRAD, UMR AGAP Institut, F-34398, Montpellier, France,

[2] AGAP Institut, Univ Montpellier, CIRAD, INRAE, Montpellier SupAgro, Montpellier, France

[3] Rice Breeding Platform, International Rice Research Institute, DAPO Box7777, Metro Manila, Philippines

[4] RiceTec Inc, PO Box 1305, Alvin, TX 77512 USA

*Corresponding author: jerome.bartholome@cirad.fr

# Abstract

Genomic prediction can be a powerful tool to achieve greater rates of genetic gain for quantitative traits if thoroughly integrated into a breeding strategy. In rice as in other crops, the interest in genomic prediction is very strong with a number of studies addressing multiple aspects of its use, ranging from the more conceptual to the more practical. In this chapter, we review the literature on rice (*Oryza sativa*) and summarize important considerations for the integration of genomic prediction in breeding programs. The irrigated breeding program at the International Rice Research Institute is used as a concrete example on which we provide data and R scripts to reproduce the analysis but also to highlight practical challenges regarding the use of predictions. The adage: "*To someone with a hammer, everything looks like a nail*" describes a common psychological pitfall that sometimes plagues the integration and application of new technologies to a discipline. We have designed this chapter to help rice breeders avoid that pitfall and appreciate the benefits and limitations of applying genomic prediction, as it is not always the best approach nor the first step to increasing the rate of genetic gain in every context.

**Key words:** rice, breeding program, genomic prediction, genomic selection, *Oryza sativa*

# 1. Introduction

The objective of every plant breeding program is to provide improved varieties that meet the needs of key stakeholders (value-chain participants from farmers up to consumers). A clear understanding of the biology and the genetics of the species combined with a targeted product concept are key elements to achieve this objective [1]. However, the genetic landscape that a breeder needs to explore to identify superior products is very large and materially exceeds the capacity of breeding programs [2]. Indeed, plant breeding can be considered as a numbers game where breeding schemes are designed to increase the probability of finding genotypes with desirable combinations of characteristics using a limited amount of resources [3]. The breeding scheme is the conceptual framework that captures all the activities that a breeder does during a breeding cycle. A single breeding cycle can be summarized in four major parts: creation, evaluation, selection, and recombination [4] and is designed to create new variation, accurately assess the performance of the breeding germplasm, and to recombine selected individuals to form an improved cohort. Evaluation is a central part of a breeding scheme which involves multiple phenotyping steps designed to estimate the heritable genetic value (or breeding value) of the selection candidates [5]. In the case of yield, usually a set of genotypes pre-selected for highly heritable traits are evaluated in multi-environment trials (MET) intended to represent the target population of environments (TPE) in which the product is expected to perform [6, 7]. These final steps of the evaluation process require significant resources and span over multiple years in a majority of plant breeding programs [3]. To overcome this limitation and increase the efficiency of breeding programs, several methodologies and tools have emerged over the last three decades due in large part to improvements in the characterization of DNA polymorphisms and computing power [8]. Among them, methods that use molecular information to infer phenotypic performance (such as marker assisted selection [9, 10] and genomic selection [11]) are important tools that allow modern breeding programs to maximize the use of their limited resources. Contrary to classical marker assisted selection, genomic prediction accounts for quantitative trait loci of both large and small effect, thus capturing a higher proportion of the genetic variance of a trait [12, 13].

The concept of genomic selection was first proposed by Meuwissen et al. [11] for animal breeding. In this simulation study, the authors predicted the genetic value based on molecular markers of juveniles without phenotypic records using the animals of the two previous generations to estimate the marker effects. They obtained high accuracies for the predicted breeding values (genomic estimated breeding values - GEBV) and concluded that this approach to increase the rate of genetic gain has potential when coupled with techniques to reduce generation intervals. Genomic selection commonly refers to the process where selection candidates, which are only genotyped, are selected based on their GEBV (genomic predictions). To achieve this, marker-phenotype relationship is first modeled using a training set (a smaller representative set of individuals that reflects as closely as possible the genetics of the individuals intended for prediction) on which phenotypic and genome-wide marker data are both generated [12, 14]. To evaluate the performance of the models, most of the time, the correlation between the predicted and observed values is calculated using a validation population whose composition depends on the validation strategy [15]. This metric is usually referred to as accuracy or predictive ability depending on which observed values predictions are compared to: breeding values or phenotypic performances, respectively.

The accelerated development of mid- and high-density genotyping technology during the 2010s led to the first report of the practical use of genomic prediction in dairy cattle [16] followed by important contributions by breeders working in agriculturally important plant species [17, 18]. Indeed, genomic prediction is now an intense field of research seeking to optimize its use and integration into both plant and animal breeding programs globally. Important advancements have been made regarding our understanding of the major factors affecting the accuracy of the GEBVs including: the effective population size of the breeding program, the heritability and genetic architecture of the target traits, the size and the composition of the training population, as well as the number, distribution, and informativeness of the markers [19]. Genomic prediction models and their implementation in software tools has also received special attention in order to efficiently leverage all information contained not only in genomic and phenotypic datasets, but also in other sources of “omics” data [20]. While the drivers of prediction accuracy are increasingly well understood, the question of how genomic prediction best integrates into an existing plant breeding strategy remains a challenge since breeding programs operate in a wide variety of contexts (target traits, species, resources, scale, etc.).

Rice (*Oryza sativa*) is a model species for molecular biology [21] and a staple food for a large part of humanity. Important gains in productivity were obtained thanks to the breeding efforts during and immediately following the green revolution [22, 23]. These improvements were realized mostly through phenotypic selection in large segregating pedigree nurseries [22, 24, 25]. The use of molecular markers was also key for the introgression of major alleles conferring resistance to biotic [26] or abiotic stress [27, 28]. The success of this strategy depended heavily on the high heritability and simple genetic architectures of the traits under selection (plant height, maturity, disease resistance, grain type) and the very large and well characterized genetic diversity of *O. sativa* [29, 30] and closely related species such as *O. glaberrima* (African rice), *O. rufipogon* or *O. nivara* [31, 32]. This may explain why the interest for implementing genomic prediction in the global rice breeding community has been delayed relative to animal breeding or breeding traditionally cross-pollinated crops like corn. During this time, it bears mentioning that some key advancements were made through population improvement via a recurrent selection strategy in Latin America [25, 33]. However, more recently, the acceleration in genetic gain for yield in other species, the decreasing costs of genotyping, and the growing importance of sustainability in rice production have contributed to an increased interest in deploying genomic prediction in rice breeding.

In this chapter, we first give an overview of the research on genomic prediction in rice with a focus on studies that make use of the strategy in a breeding program. Then we highlight important considerations for the integration of genomic prediction into a rice breeding scheme. In this second part, aspects such as identifying the entry points for genomic selection in a breeding scheme, the effective design of training populations, strategies to reduce the generation interval, and the importance of data management systems, are presented. In the third part, we take the International Rice Research Institute (IRRI) breeding program for irrigated systems as an example for the integration of genomic prediction into a product development program and provide the associated data and R scripts to run and interpret the analysis (available in supplementary information). In the last part, we present interesting progress in genomic prediction that can further help rice breeding programs to increase their efficiency. Our objective for this chapter is to provide rice breeders with a solid foundation for understanding the advantages and limitations of using genomic prediction in their breeding strategy to maximize the rate of genetic gain for relevant traits. Due to the heavy presence of inbred rice in Asia, we chose to focus the scope of this chapter to inbred Asian rice (*O. sativa*) though the specificities of applying genomic prediction to hybrid rice are addressed to a lesser extent. For another viewpoint on the importance of genomic prediction for rice breeding, we refer the reader to the book chapters from Spindel and Iwata [34] and Ahmadi *et al.* [35].

## 2. Genomic prediction works in rice

The literature on genomic prediction for crop species is very rich. With over 50 studies published since 2014, (Table 1, [36–89]), genomic prediction in rice is not an exception. We report on most of the studies published in rice (either exclusively or in concert with other species) in order to highlight the volume and diversity of the work conducted to date and their relevance for improving breeding strategies. To achieve the latter, we intentionally emphasize studies focused on integration with breeding programs, which tend to report the more practical challenges of the implementation.

## 2.1 General overview

The first studies reporting the use of genomic prediction on rice were published in 2014 (Table 1). Despite the wealth of genomic and marker resources available in rice, these studies came, surprisingly, five years after the first studies on genomic prediction (using real data) that were published in maize [90], wheat [91] or barley [92]. The breadth of genomic resources available to rice and the depth of genetic diversity that has been characterized so far has led to the discovery of many major QTL with reasonable effect sizes. While a unique and valuable resource for the rice breeding community, the heavy focus on discovery, characterization, and introgression of large effect QTL from exotic germplasm may have served to delay the transition toward genomic prediction [93]. The type of populations evaluated in these early genomic prediction studies in rice tend to reinforce that impression (Figure 1A). Indeed, among the first three studies published in 2014, two were based on the same diversity panel [94] and one on hybrids derived from a mapping population (an immortalized $F_2$ [95]). Overall, diversity panels which were, in many cases, designed for association studies [94] represented a large proportion of the studies published so far (Figure 1A). For most of these studies, the objective was methodological: understanding the impacts of population structure, integration of prior knowledge on trait genetic architecture, training set optimization, model comparison or integration of crop models without direct implication in a breeding program. Given the extent of ancestral subpopulation structure in rice, the use of diversity panels to assess genomic prediction models is likely to induce bias in the estimation of the predictive ability. Indeed if the population structure is not taken into account, most of the predictive ability can arise from the ability to predict between sub-populations and not within sub-populations [36]. Apart from studies based on diversity panels, 16 studies used breeding lines, nine studies focused on hybrids, six used a mapping population, four studies were based on synthetic populations and three used cultivars (Figure 1A).

In addition to the wide variety of populations encountered in these studies, the size of the population, the number of markers, the number of phenotypes or the number of environments used to characterize the populations were highly variable (Figure 1B). The largest population size (2,265) was achieved using publicly available data from the 3,000 rice genome project [85]. Given the limitations and difficulties surrounding the collection of high quality phenotype data, understandably most studies employed population sizes around 300 (Table 1). In cases where large populations of 1,000 or 2,000 individuals were used, the phenotyping was done in a very limited number of environments (usually 1 or 2). In fact, less than half of the studies used more than three environments for phenotypic evaluations (Figure 1B). Among the three studies having phenotypic information in 10 or more environments (year, season or location), two are based on germplasm from breeding programs [49, 64] but the datasets were unbalanced (not all individuals phenotyped in all environments or genotyped). The third study from Jarquin *et al*. [86] used the information from 51 environments in combination with days length to predict days to heading for untested genotypes. Among the wide variety of traits considered, flowering time (or maturity), plant height and grain yield were the most common. The number of markers ranged from 162 [48] to 4 million [87], with the majority of the studies using a few thousand markers (Table 1). Genotyping by sequencing and fixed SNP (single nucleotide polymorphism) arrays were the most commonly used technologies. In some cases, very high marker densities were obtained through whole genome re-sequencing at generally low coverage (1X or 2X) followed by imputation [73, 85].

Statistical methods for genomic prediction have been a central focus of many studies across all species where it's been applied. Across the 54 rice studies, 33 different methods were evaluated with the genomic best linear unbiased prediction (GBLUP) method being the most used (Figure 1C). Since this method was proposed [96], its flexibility and robustness have enabled it to quickly become a reference method for both animal and plant breeding. Similar to the traditional pedigree BLUP [97], GBLUP uses an additive relationship matrix that is based on markers instead of pedigree information. Several extensions or variations of this additive model have been proposed to account for dominance and/or epistasis [37, 53] or to use other "omics" data (transcriptome or metabolome) to estimate relatedness among individuals [80, 89]. In addition to GBLUP, RKHS (reproducing kernel Hilbert space), frequentist and Bayesian LASSO (least absolute shrinkage and selection operator), RR-BLUP (ridge regression BLUP), RF (random forest), SVM (support vector machine), PLSR (partial least squares regression), BayesB, and BayesC were the most used methods in these studies on rice (Figure 1C). Other methods from the large family of machine learning approaches, such as gradient boosting machine (GBM) or artificial neural network (ANN), were also evaluated in the context of genomic prediction with mixed results [68, 85].

The composition of the validation set, which can play an important role in determining the accuracy of predictions, was highly dependent on the validation strategy used in each study (Table 1). Sallam *et al.* [15] defined three main types of validation methods: cross validation (subset validation), inter-set validation, and progeny validation depending on the composition of the training and validation sets. The cross validation or subset validation (K-fold, leave-one-out, random, or stratified sampling) was by far the most used strategy among all of the studies that we have compiled (Fig. 1D). This validation method is very convenient because you just need to partition your data into training and validation sets to be able to estimate accuracies without an "independent" dataset (as is needed for inter-set or progeny validation). Due to its nature, cross-validation tends to overestimate the accuracy of prediction compared to more realistic validation scenarios [57, 98]. The situation becomes even more complex when multivariate models are used [99]. Another approach close to cross-validation, the HAT method [55, 100], was used in four studies. This method, based on the hat matrix of the random effects, uses the predicted residual sum of squares to estimate accuracy of prediction and works in the context of GBLUP, RKHS and Bayesian models. This method is considerably faster than cross validation as no additional model retraining is necessary [100]. The inter-set and the progeny validation methods were only used in three studies each (Fig. 1D). Considering the context of breeding programs where the integration of genomic prediction is primarily targeted to reduce cycle time, progeny validation represents a more meaningful assessment of the performance of prediction models. Indeed, in the initial concept of genomic selection, Meuwissen *et al.* [11] used progeny validation: models were built with data from generations 1001 and 1002 and the accuracies were calculated using the predicted values and the true breeding values from the generation 1003. Moreover, the decay of linkage disequilibrium occurring between markers and QTL due to the recombination in progeny generations tends to decrease the accuracy of predictions [101], but makes them more realistically interpretable in terms of applications to practical breeding scenarios. For example, Ben Hassen *et al*. [57] used progeny validation of inbred lines with a limited number of individuals and found lower predictive ability compared to cross-validation for the same traits.

## 2.2 Important findings and current limitations for genomic prediction in rice

### 2.2.1 Important findings

Table 1 provides a short summary of the main objective of each study in this review. The reader can thus be directed towards the publications that are most relevant to his or her questions.

Hereafter, we summarize important results focusing mainly on those most related to the implementation in breeding programs.

a. *Genomic prediction works in different contexts.* The most important results that arise from all studies is that the prediction of the performances based on molecular markers works. Indeed, the accuracy of GEBVs are relatively high even for traits like grain yield. Many rice breeders are concerned by the efficiency of genomic prediction but it is clearly not justified looking at the literature on rice and more specifically studies using breeding germplasm [41, 45, 53, 54, 57, 62].
b. *Prediction accuracy can be increased*. The breeders can play on different factors to increase the accuracy of predictions or to reduce the cost of implementation. Indeed, by optimizing the training set composition and evaluation, by targeting informative molecular markers (polymorphic with a medium to high minor allele frequency and spread along the genome) or by integrating additional data (historical, environmental covariates, crop model, …) better accuracies can be obtained. The size and the composition of the training set defines the strength of genetic relationship with the selection candidates which is one of the most important factors driving the accuracy. Therefore, algorithms have been developed to select the training set [39, 42, 46, 73, 78]. Concerning molecular markers, different studies show that marker density can, to some extent, be reduced without affecting the prediction accuracy. For example, Arbelaez *et al.* [67] designed a cost-effective SNP assay with only 1,000 markers selected to be informative in elite breeding material and obtained good accuracies.
c. *Models can predict offspring performance*. The initial concept of genomic selection was based on the prediction of breeding values of offspring with the objective to decrease the duration of breeding cycles [11]. The very few studies on rice that performed progeny validation [56, 57, 89] show promising results when parental information is used to predict progeny performances. However, more remains to be done in that direction since most of the increase in genetic gain related to the integration of genomic prediction is related to the reduction of breeding cycle time.

d. *Genomic prediction is efficient in the context of hybrids*: Much of the lessons learned regarding marker densities, training set identification, and model selection apply equally to hybrid and inbred breeding schemes. Hybrid programs do present unique challenges where predictions could be applied that are not applicable to other breeding schemes. Of note is the prediction of how males and females might be combined to create superior hybrid combinations. In hybrid rice there is some evidence that hybrid performance is driven by a convergence of additive genetics from the male and female lines. Incorporating non-additive parameters into the prediction doesn't seem to help [37]. While this seems reasonable, other crops have shown a significant non-additive component to hybrid performance (e.g. in corn [102, 103]). This particular conclusion was likely biased by a very narrow genetic base and very low accuracy for inter-set prediction of grain yield. There's also evidence that multi-trait models can improve prediction accuracy for low heritability traits in hybrid rice [54,81]. This is of particular importance in the hybrid context as many traits (especially cost of goods traits like hybrid seed production yield) are particularly difficult to measure early in a breeding program. A particularly unique set of correlated phenotypes associated with hybrid programs is the opportunity to measure per se performance of the inbred parents as well as hybrid performance of the same material. Using parental phenotype data combined with data on hybrid performance can improve the prediction accuracy of hybrid rice yield by about 13% [89].

e. *Modelling GxE increases prediction accuracy*. Whether it is through multi-environment genomic prediction models [56, 62] or by combining crop growth models and genomic prediction models [48, 86], several studies demonstrated the better accuracy of these approaches to predict environment specific performances. A key advantage of genomic selection over traditional phenotypic selection in the case of multi-environment models, is the ability of models to assess marker effects and marker effects by environment interactions and ultimately increase the prediction accuracy [18, 104]. With the integration of crop growth models in the genomic prediction framework, the response of the genotype to the environmental variations is modelled which allows the prediction of the performance of selection candidates for untested environments [105]. This approach is very promising for rice improvement because it takes better account of GxE. However, the routine use of crop growth models in breeding programs require a substantial investment in terms of data acquisition and analysis and thus will be interesting for specific rice systems prone to environmental constraints.

f. *Differences between genomic prediction models are marginal*. Most of the studies comparing statistical models for genomic prediction found small or no differences between them in terms of accuracy [20]. In general, none of the models is consistently better for all the traits or validation methods. GBLUP is usually used as a reference due to its simplicity, versatility to include different types of information and robustness to different trait architecture. Bayesian models (such as B-LASSO, BayesB or BayesC) or RKHS can perform better when dealing with traits influenced by large-effect genes (such as flowering time or blast resistance). The few studies that used machine learning methods (such as ANN or SVM) reported disappointing results with very variable performances even with an optimization of the parameters [68, 85]. Further work in this direction is probably needed to conclude on the interest of these methods for routine genomic prediction.

### 2.2.2 Current limitations

In spite of the number and diversity of studies, there are still some points that are not well covered in the literature on rice. Depending on the context, they can be limiting for harnessing the full potential of genomic selection.

a. *Accuracy alone is not enough to assess the effectiveness of genomic prediction*. Almost all the studies based their evaluation of genomic selection on the accuracy of the predictions. Although accuracy is an important factor to assess prediction model efficiency, it does not inform on which individuals are selected *in fine* by the different methods. The realized selection differential would probably be a better metric to compare different genomic prediction approaches as breeders jointly consider several traits to advance material, which makes the evaluation on traits separately less relevant. Finally, as rightly pointed out by Bassi *et al.* [106], the phenotype is also only a predictor of the true breeding value and has an error variance just like a GEBV.
b. *Within-family prediction accuracy is not sufficiently taken into account*. No study on rice has looked in detail into within-family prediction accuracy using multiple bi-parental families or parental information as the training set. Indeed, except for the specific case of studies using one bi-parental family, reports on within family accuracy are scarce. This is manifest as well in the hybrid literature where most papers focus on predicting specific hybrid combinations and do not attempt to estimate general combining ability among a cohort of new males or females. This is however a key point when it comes to the implementation of genomic prediction since greater within-family accuracy can help to increase the rate of genetic gain while balancing the level of inbreeding in the population. Differences between crosses are better predicted as both within and between family variations are captured by the model [107, 108].

c. *Grain quality or disease resistance traits were neglected*. No study related to nutritional value of the polished grain (zinc content, glycemic index, …) were published to date. Only one study assessed the potential of genomic prediction to help decrease the level of arsenic in the grain using breeding [72]. Regarding disease resistance, the only study from Huang *et al.* [75] reported accuracies ranging from 0.15 to 0.72 for the prediction of resistance to several isolates of Magnaporthe oryzae (blast). For disease resistance, rice geneticists focus mainly on major genes, but targeting quantitative variation is also important to address concerns like bypassing resistances. For grain nutritional value, negative correlations between traits can be better addressed using multi-trait genomic prediction.

d. *Implementation in breeding programs is secondary*. While it is clear that the underlying goal of all studies is to improve our knowledge of genomic prediction to optimize breeding strategies, few of them place their findings in a concrete case of a breeding program. For example, Spindel et al. [45] proposed to integrate genomic prediction into an irrigated rice breeding pipeline and discussed the advantages and constraints of such a scheme. However, for most of the studies working on breeding germplasm (see Table 1) this is not the case. The results therefore remain more theoretical than practical, as such analyses are important to justify investments in genomic selection and to understand potential barriers to its implementation.

## 3. Integration of genomic prediction into rice breeding programs: key aspects

Entry points for genomic selection in a rice breeding program will vary depending on the objectives of the program, the breeding strategy in place, the genetic and/or environmental constraints the breeder has to account for, and the cost of genotyping and of phenotyping the traits under selection. However, there are key prerequisites to assess before integrating a breeding program's readiness to implement genomic prediction. In the absence of essential components such as (a) clear objectives, (b) meticulous data management, (c) effective operations, (d) effective phenotyping and (e) selection based on BLUP, the application of genomic predictions is extremely limited [4]. Executing genomic prediction using breeding data or specially designed training sets is useful for establishing baseline capacity to do prediction, but integrating the technology into an existing breeding program can be a challenge. Breeding programs represent multi-year pipelines that manage overlapping cohorts of germplasm, so changing the strategy often is done step-wise so as not to disrupt the product development process. The purpose of this section is to provide guidelines regarding important elements to consider before implementing a genomic selection strategy in a rice breeding program.

## 3.1 Map the breeding strategy

The main value of genomic prediction lies in its use in decision making to efficiently select breeding material at one or several stages of the breeding scheme. Therefore, a clear understanding of the breeding strategy and its different components is the basis for an efficient integration of genomic prediction. Oftentime, the breeding scheme resides in the head of the breeder, and translating this knowledge into a structured framework is a mandatory step to carefully design alternative schemes [109]. Genomic prediction is a long-term investment for the breeding program and the direct transition to an optimal genomic selection strategy is not always possible. Therefore, a transition plan needs to be elaborated by the breeding team and experts in order to define clear steps to achieve the objectives. This aspect is usually not reported in the literature on genomic prediction as it comes down to more technical information regarding the breeding scheme. In rice, only one study placed the results in the framework of a breeding program and detailed the use of genomic prediction and its potential impacts [45]. However, as shown in wheat, this step of breeding scheme characterization is essential for the integration or the optimization of genomic selection based on the knowledge acquired during the last years [106, 110].

Optimal genomic selection schemes are usually not simple evolutions of the current breeding scheme. The majority of conventional breeding schemes in rice, and self-pollinated crops in general, rely on pedigree breeding [25] but genomic selection is best suited to recurrent selection schemes based on elite by elite crosses to improve complex traits. Indeed, a well structured breeding program where the elite germplasm has been clearly identified and with a small effective population size ($Ne \approx 40$) is more likely to benefit from the use of genomic prediction due to higher linkage disequilibrium between markers and QTL, low or absence of population structure and higher relatedness among genotypes. In addition, several major changes are needed to fully leverage genomic predictions: reduce cycle time, build a training set, store / use phenotypic and genotypic data, reallocate budget and staff [106, 111]. Understanding the interconnections between these changes and how they will impact the sequence of current operations allow to anticipate potential obstacles.

*Key recommendations*:

a. Define clearly the current breeding strategy and its objectives.
b. Plan the integration of genomic prediction as a long-term investment with a clear roadmap.
c. Use recurrent selection in elite population to maximize the potential of genomic prediction

## 3.2 Reduce the cycle time

An interesting aspect of genomic selection is that it has led to a greater focus on the fundamentals of breeding in the plant breeding community [112]. The concept of response to selection captured in the Breeder's equation is perhaps the best example [4, 109]. Among the parameters of the equation, the generation interval (or cycle time) is the easiest to understand and to play with. As highlighted by Meuwissen *et al.* [11] in their seminal paper, the use of genomic predictions can greatly increase the rate of genetic gain by reducing cycle time: "*It was concluded that selection on genetic values predicted from markers could substantially increase the rate of genetic gain in animals and plants, especially if combined with reproductive techniques to shorten the generation interval.*" This conclusion was confirmed fifteen years later by the first report of the impact of genomic selection on the rate of genetic gain in dairy cattle [113]. The authors found a dramatic reduction in the generation interval related to a sharp increase in the rate of genetic gain from yield traits (50 - 100%). In plant breeding, methods to reduce cycle time (independently from the use of genomic selection) have been studied for several decades now [114, 115] . Rapid generation advance (RGA) or double haploids are probably the most common in crop species, even if more modern approaches have been proposed lately [116, 117]. In rice, RGA has regained interest recently as it is a cost-efficient way to quickly fix material (typically from F2 to F6 in one year) for its evaluation in replicated trials [118]. This can be realized in greenhouses, screenhouses or in the field depending on the resources available. For breeders working on a classical pedigree breeding scheme, the use of RGA could be a first step toward the implementation of genomic selection [119]. For breeding programs already implementing RGA or similar methods to reduce cycle time, genomic selection can further help to shorten the breeding cycle. However, this requires a genomic prediction model that can efficiently predict the genetic value of the next generation (progeny). Therefore, a training set based on material from one or several previous cycles has to be constituted before implementing this type of scheme. This is also the case for more aggressive strategies based on recurrent selection that aim at recombining non-fixed material ($S_0$) selected based on predicted values only. In that type of scheme, the population improvement part is partially decoupled from the product development part which allows a 1-year or even shorter breeding cycle [120, 121]. For the moment, only simulation studies have reported this type of scheme since several technical challenges have to be solved before implementation. Indeed, a drastic reduction of breeding cycle time can lead to overlapping activities between different cycles during the transition period that may disrupt ongoing cycles or increase substantially the workload.

*Key recommendations:*

a. Use genomic prediction in conjunction with robust methods to produce inbred lines (e.g. rapid generation advance) to effectively reduce cycle time.
b. Take into account technical constraint associated with cycle time reduction into the genomic prediction roadmap.

## 3.3 Design the training set

Once the entry point of genomic prediction in the breeding scheme has been defined, the design of the training set is the first step toward the implementation of genomic selection. Three major choices have to be made regarding the training set: its composition and size, its phenotyping and its genotyping. The breeder must find a balance between these three aspects in order to optimize the training set according to available resources. A simple way for most breeding programs to get started is to begin genotyping every line that enters the yield trial. From there, those datasets can be empirically optimized to increase prediction accuracy.

It is well known that the accuracy increases with the size of the training set. Theoretical [122–124] and empirical studies [65, 125, 126] suggest that the training set size should be maximized when dealing with complex traits. However, large training sets are not always feasible mainly due to genotyping and phenotyping costs. Several methods were developed to optimize the training set composition in order to achieve high accuracies while maintaining the size to a manageable number [39, 42, 46, 69, 73, 78, 127–129]. All of these methods use the additive genetic relationships (usually based on marker data) to optimally sample a set of representative genotypes. A key aspect of the optimization of the training set is the definition of the predicted set (selection candidates). Indeed, close genetic relationships between the training set and the selection candidates are key to maximize prediction accuracy [130, 131]. Therefore, most of the optimization methods are jointly considering the genotypes that will compose the training and the predicted sets to either directly compute criteria based on relatedness (the average of the relationship coefficients between the training set and the predicted set [128, 132]) or to estimate criteria based on mixed model theory (the prediction error variance, the coefficient of determination or the expected accuracy [39, 78, 127]). In the cases where the training and the predicted sets come from the same population (e.g. selection candidates from the same cohort) or the information on the predicted individuals is not yet available (e.g. offspring), optimization methods have been developed to minimize the genetic relationships between individuals of the training set [46, 73]. Depending on the availability of data and the prediction objectives, the breeder can choose among these optimization methods to shape the training set and update it when selection candidates from a new cycle need to be predicted.

The optimization of the composition of the training set has to be done in conjunction with the phenotyping strategy. In most cases, the selection candidates that will be used to update the prediction model are evaluated for key traits in MET to estimate G×E. Since the total number of plots available for the evaluation is almost fixed, the breeder needs to balance the population size with the level of replication (within and across environments). Classically, the level of replication increases during the breeding cycle to dedicate more resources to a smaller number of more promising lines in the final stages. In the context of genomic selection where the evaluation unit being the alleles instead of the individuals, increasing the size of the training set while decreasing the level of replication tends to increase the accuracy of prediction [133, 134]. The typical size of a training population (150 - 300) to be phenotyped in a classical fully replicated experiment can therefore be multiplied by 1.5 to 3 with sparse testing. However, it is advisable to have a sufficient level of replication within and across environments to: i) maintain repeatability, especially for low heritability traits, ii) assess the level of G×E and iii) avoid model convergence issues with too few replicates. The limitation of replication using sparse testing approaches can also be a good opportunity when the seed availability is a constraint.

Finally, the technology used to genotype the training and predicted sets needs to be carefully considered in order to efficiently capture distinct QTL alleles as well as general relatedness in the population. Several factors come into play when choosing or developing the appropriate genotyping technology: cost, type of markers, density, informativeness in the target population, reproducibility rate, etc. In the case of applying genomic prediction, a good characterization of the genetic diversity managed by the breeding program is essential to determine the marker density needed to achieve an optimal prediction accuracy. It has been shown using both deterministic [13, 135] and stochastic [136] simulations that the marker density has to increase when the effective population size increases to maintain the accuracy [135–137]. However most empirical studies in rice found that the accuracy reaches a plateau when the marker density goes beyond 2 to 5 markers per centiMorgan for breeding programs with an effective population size lower than 50.

*Key recommendations*:

a. Maximize the relatedness between the training and the predicted sets where possible.
b. Use sparse testing for phenotyping in order to balance the size of the training set and the level of available resources.
c. Avoid using a training set from one breeding pipeline in order to predict the candidates from another breeding pipeline.

## 3.4 Generate and integrate good quality data

As highlighted before, data acquisition and management are essential components of a breeding program. All advancement decisions are made based on recorded data from multiple sources (field, laboratory, service provider, etc…). Careful data management from the seed to the phenotype and/or to the genotype have to be in place to ensure accuracy. The use of digital data collection tools are a key way to reduce as much as possible errors that can be perpetuated during the data collection process. Concerningly, it has been demonstrated with simulated data that even a small percentage of severe errors (0.1% or 1%) in phenotypic records can severely reduce the response to selection [138]. Similar conclusions were also found when errors are present in the pedigree records [139]. Besides accurate data, robust and appropriately designed analysis pipelines are needed to curate the data and turn it into interpretable intelligence. Genomic prediction adds an additional layer of complexity compared to traditional marker assisted selection in that it can require the integration of different types of data (phenotypes, genotypes, pedigree, and/or weather data) collected over several years to be useful. Consistency of data type and format and the stability of data structures over time is a key aspect to leveraging the full power of historical breeding data to train and continuously update genomic prediction models [140].

To help the breeders with data management, software solutions such as the Breeding Management System (https://bmspro.io), Breeding4Results (B4R)

(https://riceinfo.atlassian.net/wiki/spaces/ABOUT/pages/326172737/Breeding4Results+B4R), Breedbase (https://breedbase.org), or GOBii Genomic Data Management (https://gobiiproject.atlassian.net/wiki/spaces/GD/overview) are available and used in different public organizations. Despite the significant efforts to develop analysis pipelines (like the RiceGalaxy, https://galaxy.irri.org, [141]) and the Breeding API project (https://brapi.org) designed to enable interoperability among plant breeding databases, no efficient end to end solution is publicly available to perform genomic prediction in the context of an applied breeding program. Indeed, several limitations are present among available software for implementing genomic prediction, including a lack of direct linkages between genotypic and phenotypic data, limited multi-environment or multi-trait analytical capability, no possibility to integrate dominance or epistasis effects into a prediction model, and no meaningful integration of weather data into an analytical pipeline. The majority of public breeding programs therefore extract the phenotypic and genotypic data from their respective data management software and use ad hoc analysis pipelines to run genomic prediction models. Hopefully, projects such as the Breeding API or the Enterprise Breeding System (https://ebs.excellenceinbreeding.org) will offer these possibilities in the near future within a coherent framework designed to enable applied breeding programs.

*Key recommendations*:

a. Use digital data collection systems where possible.
b. Work with data management systems and efficient analysis routines for genomic prediction (GBLUP, RR-BLUP).
c. Use consistent genotypic and phenotypic data structures over years to facilitate data integration.

## 3.5 Take into account the costs

The integration of genomic selection in a breeding program is a long-term investment that must translate into a better rate of genetic gain to be worth implementing. Even if the advantages of using genomic selection are clear, the optimal breeding scheme relative to genetic, operational, and cost constraints is not easy to identify. After setting a vision for what's optimal, the need to convert to this new strategy in a budget friendly way is probably the most important limitation for the strengthening of modern breeding programs. Nevertheless, there are several levers that can be used to liberate resources in a program aiming to fully deploy genomic selection.

The first levers are related to phenotyping. Thanks to genomic prediction, some phenotyping steps can be reduced or even eliminated saving the related costs *de facto*. Indeed, this is one of the main advantages of genomic prediction which, with the right data structures in place, allows for both a reduction of cycle time and phenotyping costs [111]. The costs of phenotyping and the potential to replace a phenotyping activity with a prediction should be carefully evaluated when planning the integration of genomic prediction as it may sometimes require a modification of the breeding scheme. One key example of this is the cost savings incurred when transitioning from traditional pedigree breeding program where the selection that occurs during the fixation steps (F2 to F5) can be delayed until after inbred lines have been extracted by substituting a field based pedigree nursery with a much cheaper and faster SSD based RGA method. The cost savings made at this level can easily cover the cost of genotyping since advancing material through RGA is much less expensive (around 1 US dollar per F5/F6 lines) [119]. Organizations must however look to multi-year budgeting strategies to accommodate the fixed costs that may be incurred if existing greenhouse facilities cannot be leveraged for this activity. Initial capital investments can often be paid for by reduced operational costs over several years. Furthermore, organizations must factor in the additional funding that could be generated due to the increase in genetic gain that will accompany a shortening of the breeding cycle and an improvement in selection accuracy.

Another direct way to recover costs is by using genomic prediction to reduce the volume of an expensive phenotyping exercise [73, 142]. This can be done either by selectively phenotyping a carefully chosen subset of a trial for expensive traits like grain biochemistry or other post-harvest traits and using the cost savings to pay for DNA fingerprinting. Additionally, developing an index of high throughput correlated traits that may be less expensive to measure or offer higher throughput compared to the target trait can decrease the cost of phenotyping and offer similar accuracy. In that context, multi-trait genomic prediction offers an ideal framework to integrate correlated traits to maximize prediction accuracy [143].

The second levers are related to genotyping. In a crop breeding program, the choice of the genotyping technology to characterize the breeding germplasm (training and prediction sets) is mostly driven by the cost of genotyping per sample (and not really well captured by the cost per data point) [144]. Indeed, the cost per sample with available tools (genotyping-by-sequencing or fixed SNP arrays) is often too high to be used routinely in a public breeding program. In small to medium size breeding programs, the cost per sample has to be around 10 US dollars or less in order to assess a sufficient number of individuals. In that price bracket, the number of loci that can be currently targeted is around 1,000,- 5,000 SNPs. One option to keep costs down in the long-term is to design a custom genotyping assay with SNPs selected to be specifically informative in the target breeding population. This would be a cheaper option than GBS or public fixed arrays and allow for higher density of information content in the genotype dataset. A custom SNP panel has the additional benefit of potentially surveying specific trait markers of relevance to a breeding program in addition to the genome-wide markers included in the set; thus allowing for more extensive QTL profiling of lines for known alleles that are not necessarily prioritized for MAS. In fact, depending on the capability of the genotyping service provider, it is not unreasonable to save sampling and DNA extraction costs by combining MAS and fingerprinting such that the cohort is screened with a few markers intended for MAS, then to have the DNA from selected lines re-arrayed into a new plate for genome-wide fingerprinting.

It is also possible to achieve low genotyping cost by using low-coverage genotyping-by-sequencing [145]. Given the limitation of genotyping-by-sequencing when the sequencing depth is lowered (high rate of missing data, high error rate for heterozygous loci), this approach won't capture heterozygous loci efficiently and must be used for genotyping fixed lines, coupled with an efficient imputation framework based on high quality sequence data of ancestral lines in the pedigree. This therefore requires expertise in bioinformatics and access to high performance computing resources.

*Key recommendations*:

a. Consider reducing the number of phenotyping steps, only phenotyping a subset of a trial, or using cheaper or higher throughput correlated traits.
b. Design a genotyping platform with a set of markers selected specifically for the germplasm managed in the breeding program and deploy it at a service provider.

# 4. An example on IRRI breeding program for irrigated systems

Here, we give a practical example of the integration and use of genomic predictions in an active rice breeding program. The recently re-designed breeding program for irrigated systems at IRRI offers an ideal context to understand the key elements of an applied breeding program using genomic predictions [146, 147]. Indeed, with its global mandate of Southeast Asia, South Asia and Eastern Africa as the main areas of intervention, it represents the direct derivation of the early breeding efforts that resulted in the Green Revolution in Asia. As such, it is the best representation possible of an effort to produce materials that combine high yield potential and adaptation to diverse environmental conditions.

## 4.1 The transition from pedigree breeding to recurrent genomic selection

The applications of genomic selection to the IRRI breeding program came in two broad categories: within cohort predictions (full and half sibs predicting other full and half sibs) to optimize our testing strategy and across cohort predictions (grandmothers and mothers predicting daughters and granddaughters) to accelerate our breeding cycles, both of which required changes to the breeding strategy. First and foremost, both applications required the cost-effective deployment of a genotyping technology that allowed for the routine fingerprinting of the breeding material. This marker set (known as the 1k-RiCA amplicon panel [67]) had recently been developed and populated with markers that were specifically informative in our germplasm. Publicly available fixed array genotyping technology would not have served this purpose well as many of the markers on these arrays are chosen to differentiate germplasm globally [148] and were often very expensive with relatively few (or worse, biased) polymorphisms.

With the marker panel in place and deployed at a service provider, in the immediate term, the most useful application of genomic selection was to allow for selections to be made based on performance in the target environments rather than depending on a correlated response to selection with Philippine environments (where IRRI's headquarters are located). The program as it is currently resourced generates a stage 1 yield trial of approximately 2,000 new lines each year. As all of IRRI's yield trials are conducted by national agricultural research partners, the ability to test 2,000 lines in multi-location yield trials in Africa, South Asia, and Southeast Asia was extremely limited. Up to this point, the early generation breeding material was selected based on performance in the Philippines and a small number of advanced lines were sent to the regional locations for testing and evaluation (Figure 2). Genomic selection using full and half sibs was employed to enable direct selection based on the target environment and avoid needing to rely on indirect selection. By selecting an optimized subset of the cohort and sending it to be tested in the region of interest, phenotype data from the specific region of interest could be used to predict the performance of the remaining cohort in that region. In this way, the entire cohort is tested somewhere, but no individual is tested everywhere, and thus an advancement of superior lines can be sent to partners that is tailored to their unique conditions. To do this, however, required that funds be identified to fingerprint the full cohort of about 2,000 new lines every year. In order to make this form of genomic selection cost neutral, it was noted that the testing strategy in the Philippines was testing lines for three years (Figure 2, former scheme). By eliminating the middle testing phase and selecting a region specific set of lines for advance testing, sufficient funds were recovered to cover the cost of fingerprinting.

The genomic prediction application with more long-term value to the program was to enable across cohort predictions so that superior lines in each region could be recycled back into the breeding pipeline prior to regional testing, and thus accelerating the breeding cycle (Figure 2, future scheme). This kind of prediction however requires a more robust, multi-year dataset consisting of regional phenotype data on ancestral lines, as phenotype data from full and half-sibs of the emergent candidates would be unavailable at the time the prediction needs to be made. With the first application of genomic prediction in place, the program is now well positioned to begin generating multi-year datasets with region specific phenotypic observations needed to predict new parents. However, to make this kind of prediction possible, a more directed manipulation of the crossing strategy needed to be implemented. The most important decision a breeder makes is selecting and crossing parents on the basis of breeding values for relevant traits. As this metric was not routinely calculated at IRRI, our first step was to gather our historical data together into a single model and generate the best estimates possible for breeding values and reliabilities for yield, maturity, and plant height. Breeding values for other important traits such as grain quality, disease resistance, and other agronomic traits were not collected routinely enough or at enough locations to provide meaningful estimates of breeding value. This process was substantially accelerated due to the efforts made to migrate data into the B4R data management system. As DNA fingerprint data was not available on the vast majority of our historical lines, pedigree data stored in the genealogy management system was used to estimate relatedness coefficients. This multi-year evaluation of our historical data permitted the identification of a unique core set of lines with high and reliable breeding values for yield, which would form the basis of further breeding and germplasm characterization efforts. Once identified, this set of high breeding value lines were fingerprinted and that data was then used to estimate the effective population size and used to estimate the frequencies of major genes for other traits (such as amylose content or resistance to blast). These metrics would be used to guide selection strategies among the progeny and evaluate the risk/benefit of introducing new genetics into the program.

This step, while not specifically motivated by genomic selection, was critically important because along with the development and characterization of the core germplasm came a commitment from the program to primarily cross within this new gene pool to drive genetic gain. This relatedness across generations (and aversion to frequent introduction of new germplasm into the program) creates genetic continuity over multi-generational cohorts that enables the ability to use phenotype data from ancestors to predict the performance of newly created descendants. Corresponding with that relatedness was the development of business rules for crossing and population development. These rules ensure that new crosses generated by the breeding program maximized genetic variation in the next generation to the extent possible. They also allowed for sufficient numbers of full and half-siblings in each cohort to be generated, from which predictive power could be obtained. Among these, business rules included a commitment to cross with lines from the most recent cohorts whenever possible (rather than older released lines), preventing the use of any one line in more than 10% of the crosses to avoid bottleneck the variation, the complete avoidance of sub-lining so that each F2 plant generates a unique F6 line, and ensuring that sufficient new fixed lines from each cross were entered into the stage 1 yield trial such that there was a reasonable probability of identifying a new line that was at least one standard deviation better than the average yield of the cross.

With these two applications of genomic prediction underway, the program went from a long-cycle pedigree nursery to a rapid-cycle genomics enabled breeding strategy. This strategy involved making crosses and setting population size targets according to predefined business rules, generating new lines through RGA approaches, employing MAS after line fixation, and using bulk harvests of the selected head rows to create seed for shipping to regional locations for testing. Predictions of the entire cohort across all regions would ensure that every line had either an observation or a prediction in every region, from which a core set of superior region specific lines were identified and shipped to partners for stage 2 yield trial evaluation and testing. As data accumulated in the regions on cohorts of lines, and as the progeny and grand-progeny of the original core set of lines begin to fill the pipeline, the capacity for predicting regional performance across cohorts will grow until sufficient data becomes available to allow for the identification of new parents prior to stage 1 yield testing.

## 4.2 Description of the breeding schemes and integrating genomic prediction

The mapping of the breeding scheme is a key component for the optimal use of breeding program resources and to understand where the entry points for genomic selection could be placed. The current breeding strategy summarized in Figure 2 was initiated in 2017 at IRRI in order to reduce cycle time and optimize multi-environment evaluations thanks to the introduction of genomic prediction. In this strategy, most of the activities take place at IRRI headquarters in the Philippines. The first year, the crosses (80-100) are made and the F1 plants are validated using dedicated SNP markers. The second year the segregating families go through SSD from F2 to F6 via RGA. At that stage, 7,500 to 10,000 lines are advanced: this corresponds to 200 - 400 lines per cross. Population sizes for each cross are determined based on the anticipated segregation of major genes. The third year, the lines are evaluated in the field in panicle rows for seed increase and for the evaluation of uniformity, plant architecture, and maturity. At the same time, the lines are genotyped for marker assisted selection for major loci prioritized for each breeding pipeline. These include the waxy gene for amylose content and a number of disease resistance genes for major pests and disease (blast, bacterial leaf blight, ...) [10]. The second season of the third year is dedicated to the preparation of the seeds to be shipped in the regions. The fourth year, the lines advanced based on MAS and head row selection (1,500 - 2,000) are genotyped using a low density platform with less than 1,000 SNP markers [67]. The same lines are also evaluated in the first stage yield trial at IRRI headquarters in the Philippines. In parallel, a subset of the cohort (250 - 300 lines) is sent to the regional partners in South Asia and Eastern Africa for multi-environment evaluation of key agronomic traits (plant height, time to flowering, grain yield). This subset (training set) is used to build the genomic prediction model that is later used to select an advanced class of superior lines among the entire cohort. Since no historical data were available for building reliable genomic prediction models, the integration of genomic prediction in this scheme relies on the use of half-sibs or full sib-sibs to maximize the accuracy with highly related training and predicted sets [142, 149]. The genomic prediction models are used to select parental lines for the following cycle and to select promising lines (30 - 40) for the second stage yield trial that are conducted in the fifth year of the breeding scheme. The best performing lines at the end of this stage can then go through advance testing in the national variety release system or can be used by partners in the regions in their breeding program to enrich their gene pools.

In this strategy, the breeding cycle spans over five years with the recycling of advanced lines as parents occurring during the fourth year (Figure 2). Compared to previous breeding schemes that were in place at IRRI, the cycle time is shortened by 2 years [147]. Reduction of cycle time is a key factor to increase the rate of genetic gain [109]. In this scheme, one of the major tools for cycle reduction is RGA. This approach, known for a long time [150, 151], was optimized in 2013 and implemented at large scale at IRRI in 2014 [118]. Currently, genomic prediction is not used to decrease cycle time and is mainly used to increase the intensity and accuracy of selection in regional environments, especially for yield. The main reason for this is the lack of historical data in the breeding program suitable for genomic prediction. Indeed very few breeding lines have been consistently genotyped and phenotyped to build a reliable database. Therefore, the current phase is a transition phase where the data currently generated feeds a database that will be used to predict the performance of future progeny (across cohort predictions). This is highlighted in Figure 2 as the future scheme. This ability to directly predict the performance of selection candidates before evaluating them in the field will enable us to decrease the cycle time by two additional years resulting in a two-year breeding cycle. However, this comes with operational challenges such as: ensuring four generations per year in a stable manner during RGA, production of enough seed at the end of the RGA to enable multi-environment trials, and navigating the import/export process quickly enough to ensure the seed arrives to the partners in time for planting in the main season.

## 4.3 A Practical example of the analytical pipeline

In this section, we present the analysis pipeline that we currently use at IRRI to perform genomic selection. This corresponds to the activities mapped to the fourth year of the current breeding strategy (first stage yield trial, Figure 2). The analysis pipeline is divided into three main steps (Figure 3):

a. *The selection of the training set.* This step is based on SNP markers specifically chosen to be informative in the elite germplasm used in the breeding program [67] and the optimization method of Akdemir et al. [39] that minimizes the prediction error variance (PEV) in the predicted set.

*b. The single trial analysis.* In this step, phenotypic data (plant height, days to flowering, and grain yield) are measured on the training set in several regional locations, which are analyzed separately to assess the quality of the data at each location and estimate spatial adjustments to genotypic values with a mixed model, taking the experimental design into consideration.

*c. The genomic prediction analysis.* In this last step, a GBLUP model trained with the genotypic and phenotypic data from the training set is used to predict genomic estimated breeding values (GEBVs) for all the untested lines.

To illustrate the analysis pipeline, real data from the IRRI breeding program for irrigated systems is used as an example. The analyses were conducted within the R environment and utilized the R packages *asreml* (under license) or *sommer* (freely available) for mixed model analyses and functions developed specifically for the analysis pipeline and from the literature. We have opted to give the user the possibility to choose between *asreml* and *sommer* according to his preferences. All the R scripts are provided in the Additional file 1.

#### 4.3.1 Selection of the training set

In the current breeding scheme the genomic prediction is used for within cohort predictions. In order to identify the best subset (training set) to be phenotyped in regional MET, we use an optimization method based on mixed model theory that minimizes the prediction error variance [39]. This method available in the R package STPGA (for Selection of Training Populations by Genetic Algorithm) requires the genomic relationship matrix (G matrix) as an input. In the example, the entire cohort of 1,722 lines is genotyped with 1,079 SNP markers. We use the rrBLUP package to compute the G matrix based on the genotypic matrix (geno_data) containing marker information coded as [-1, 0, 1]. The G matrix is then used as a parameter for the *OptiTS* function along with the desired size of the training set (sTS = 300) and the number of replicates (rep = 5). The number of replicates allows the selection of the individuals most represented in the different runs to be included in the training set in order to avoid suboptimal solutions from the genetic algorithm [39]. To evaluate the representativeness of the training set compared to the entire cohort, the individuals are plotted using the two first principal components from the G matrix (Figure 4).

#### 4.3.2 Single trial analysis

Once the training set is identified, it is sent to regional partners to be evaluated in MET. For this case study, actual trial data from five different locations in Bangladesh were used. These trials were conducted in the 2020 dry season (Jan - May). Each trial comprises 362 breeding lines of which 299 are training set lines, and the rest are advanced lines from the previous cohort and check varieties. All the trials used a partially replicated design with 20% of lines replicated. Three traits are used in this example: plant height (cm), days to flowering, and grain yield (t/ha). The trial data is uploaded into the B4R database, which has been adopted by IRRI for managing all breeding trial data. The exported data from the B4R database for each location is used to perform individual single trial analyses (pheno_data object). The objective of this step is to remove potential error in the dataset and to adjust from spatial variation using the experimental design. The following mixed model (*asreml* or *sommer*) is used to obtain the BLUP and deregressed BLUP for each line:

```
model <- asreml( fixed = trait ~ 1 ,
                 random = ~ DESIGN_X + DESIGN_Y + GID,
                 na.action = na.method(x = "include"),
                 data = dataset)

model <- sommer::mmer(fixed = trait ~ 1,
                      random = ~ DESIGN_X + DESIGN_Y + GID,
                      rcov = ~ units,
                      data = dataset,
                      verbose = FALSE)
```

The variable DESIGN_X and DESIGN_Y represent the coordinates of the plots within the field. The variable GID represents the ID of the genotypes. The BLUP and deregressed BLUP values are then calculated. The single trial analysis is embedded in a function called *single_trial_asreml* or *single_trial_sommer* that takes the formatted phenotypic raw data as an input and returns a data frame with several variables including: location, trait, genotype ID, BLUP, de-regressed BLUP, and repeatability ($H^2$). The function is then used for all locations and traits to run the model and retrieve the BLUPs (Figure 5A).

#### 4.3.3 Genomic predictions

The deregressed BLUP value of the training set lines from the single trial analysis and the genome-wide marker genotype data of the entire cohort (training set and predicted set) consisting of 1,722 lines are used in the genomic prediction model. The genome-wide marker data is used to construct the additive relationship matrix with the *sommer* package. The inverse of the additive relation matrix is then constructed in the case where *asreml* is used the GBLUP analysis. The GEBV for each line is computed using the GBLUP model where the regressed-BLUP from each location is the response variable, location as fixed effect, the breeding line (gid) and inverse of the G-matrix (invG) are used as the random effects.

```
model <- asreml(fixed = trait ~ 1 + location,
                 random = ~ vm(gid, invG),
                 data = dataset)

model <- sommer::mmer( fixed = trait ~ 1 + location,
                        random = ~ vs(gid, Gu = G),
                        rcov = ~ vs(units),
                        data = dataset,
                        verbose = FALSE)
```

Similarly to the single location analysis, this model is embedded in a function (*gblup_asreml* or *gblup_sommer*) with two parameters: the first is the output from the single location analysis and the second is the inverse of the G matrix. The output of the function is a table containing the GEBV on the entire cohort (Figure 5B). The GEBV values are then combined with trait marker information and used by the breeder for selecting lines for advanced testing and, also, selecting parents for the next breeding cycle.

# 5. Other applications of genomic prediction for rice improvement

In the previous parts of the chapter, we saw that genomic selection requires both methodological research and a carefully designed breeding program to be implemented efficiently. In this last part, we present ongoing developments regarding the use of genomic predictions for rice improvement. We think it is important for breeders to be aware of upcoming approaches and tools to be ready when they are mature enough to be integrated in breeding programs when appropriate.

## 5.1 Characterization of genetic diversity for pre-breeding

The characterization and the use of genetic diversity is important to meet long-term breeding objectives and maintain the adaptive potential of the breeding populations [152]. In the case of recurrent selection in elite germplasm, the addition of new material threatens the genetic gain in the short term by diluting the impact of high value alleles carefully accumulated through successive cycles of selection. However, in the long-term, the loss of genetic diversity due to selection but also to negative or neutral linkage drag or genetic drift can be compensated by careful introduction of genetic variation into the elite pool [153]. The identification of the best accessions for particular breeding objectives is laborious, as it requires an accurate phenotyping of a large number of diverse lines that often mask valuable haplotypes in low breeding value backgrounds. In this context, genomic prediction can be used to identify superior accessions in germplasm collections and be applied to pre-breeding, which aims to identify high-potential genotypes among a large number of accessions [154–156]. In rice, the availability of large genomic resources such as the 3,000 rice genomes [30] or the high-density rice array panel [157] offer a unique opportunity to use genomic prediction to target valuable genotypes relative to the breeding objectives.

## 5.2 Definition of heterotic groups for hybrid breeding

In hybrid breeding, heterotic groups are usually needed to optimally use the heterosis within a species [158]. To this end, hybrid selection causes the germplasm to become structured into genetically distinct groups that display superior hybrid performance when individuals from complementary groups are crossed. Contrary to other major crops (e.g. corn [159), heterotic groups in rice are defined largely according to complementarity with a particular sterility system and not according to gene pools defined by complimentary heterotic potential. This is further complicated in rice due to the strong population structure that characterizes rice diversity being confused as heterotic differentiation of complementary gene pools  [29, 30]. Efforts to coerces ancestral subpopulations into heterotic groups, as in the case of the two major types (*indica* and *japonica*), have limitations due to sterility, contrasting adaptations, and very different distributions of major grain quality parameters[160]. Further research is required to identify natural patterns of heterosis [161], and in some cases genomic prediction can assist this exploration. Recently, the use of predictions to define heterotic pools based on complementary yield performance has been proposed in rice [162]. In this study based on real data, the authors applied the approach developed by Zhao et al. [163] to detect heterotic patterns for yield by combining the predicted performances of all unique single-cross hybrids with a simulated annealing algorithm with different group sizes.

## 5.3 Integration of high-throughput phenotyping and environmental information

The significant progress made with genomics in breeding programs has reinforced the idea that phenotyping is still a bottleneck for genetic improvement [164]. This may seem paradoxical since one of the advantages of genomic selection lies in the reduction of some phenotyping steps. However, accurate field phenotyping for important traits (e.g. grain yield) in METs is even more important to efficiently train the prediction model and capture G×E. In addition, selection for more expensive or difficult traits (drought resistance, lodging tolerance, grain quality, etc... ) can be integrated earlier in the breeding scheme thanks to genomic prediction and therefore increase the selection intensity. These observations have led to an ever increasing interest in high-throughput phenotyping methods [165, 166]. Several tools (RGB and multispectral cameras, thermal sensor, etc.) and platforms (phenomobiles, unmanned aerial vehicles, etc.) are available for field and laboratory phenotyping with a wide range of applications. When integrated in a genomic prediction model, high-throughput phenotypic data can substantially increase the prediction accuracy [167, 168]. In the case of phenomic selection, high-throughput near-infrared spectroscopy data can even replace genotypic data and offer similar accuracy [169, 170]. However, to be useful in a breeding context, the large quantity of data generated by the high-throughput phenotyping techniques needs to be stored in a data management system, properly vetted relative to the costs and selection accuracies available from manual phenotypes, and associated with correct genotype data if it is to improve the decision making process. Although tools and analysis pipelines have evolved in recent years, there are still important constraints to the routine use of these approaches: the acquisition of multi-environment field data and not just data from a central research station, the availability of data management systems that can handle large time-series datasets, and the initial cost of related equipment. It's expected as the technologies and regulations mature, that dedicated companies offering high-throughput phenotyping services will emerge, much like has been the case with genotyping.

In addition to high-throughput phenotyping, a better characterization of environmental factors affecting the performance crop plants will enhance our ability to explain non-genetic sources of variation. Such "envirotyping" is an area of active research that shows great promise [171]. To become truly useful technologies that permit the high-throughput collection of envirotype data in real time need to continue to mature as well as data management and analytical strategies for extracting intelligence from these datasets.

## 6. Conclusion: a point a view of a rice breeder

Based on the literature in rice and in other species, the ability to do genomic prediction and the value of applying genomic selection to rice breeding programs is beyond question. The capacity to estimate the prediction values and the key datasets and models that underlie the estimation of GEBVs is also very well understood. The marker resources and phenotyping capacity in rice are present and available at this point to even the most remote breeding organizations. Furthermore, the rules that describe how quantitative trait variation is inherited in populations is well understood and it would seem the infinitesimal model applies to quantitative traits in rice in most cases. What remains to capture the full value of this technology is the reorientation of rice breeding programs around a short-cycle recurrent selection strategy within a defined gene pool. During that transition, genomic prediction can additionally be helpful for improving selection within cohorts and save money on field evaluation. As a result, generating genotype data or building an analytical pipeline is often not the starting point for implementation of genomic selection in most programs. Clear business rules for data collection and management, clearly defined best practices for parental selection and a commitment to work within elite gene pools must come first. Second to these foundational activities, breeding programs must standardize and systemetize their operations in such a way that resources are optimized, workflows are clear, and breeders are not spending inordinate amounts of time managing logistics. Field work needs to focus more on data quality and data collection, reserving selection decisions for after data has been collected, analyzed, and interpreted. Marker systems for routine genotyping are also necessary, but must be developed such that the genotype data is specifically informative to the breeding germplasm of interest.

The public rice literature to date has largely focused on questions related to if predictions work in rice or how to optimize prediction accuracy. Very few rice publications address how predictions can be practically applied to enhanced rates of genetic gain. As a result, in an attempt to modernize many breeders get stuck in ‘proof of concept purgatory’ by trying to replicate analyses done by others. Breeders seeking to improve their strategy would instead be benefited from considering whether the appropriate foundations are laid in their programs and then considering carefully what the entry points for prediction are in their stated breeding strategy. Commercial breeding programs may have the advantage of having the freedom to invest resources in additional capital or operational expenditures up front in order to capture value in the long term. However as budgets are often tight, fixed, or subject to congressional approval for publicly funded programs, cost saving adjustments to the breeding strategy (such as applying a sparse testing design or implementing rapid generation advance for line fixation) may liberate resources in the short term which can be applied to laying the proper foundations for a fully genomic prediction-enabled breeding strategy.

# Acknowledgements

The authors are grateful to Adam Famoso and Flavio Breseghello for their valuable comments and comprehensive review of the chapter. We would also like to thank the irrigated rice team at IRRI: Rose Imee Zhella Morantte, Vitaliano Lopena, Holden Verdeprado and Juan David Arbelaez, for their help with data acquisition and management regarding the example provided on IRRI breeding program. We thank the IRRI Bangladesh team and in particluar Rafiqul M.Islam as well as our partners in Bangladesh for their support in obtaining phenotypic data for the training set presented in the exemple.

# Funding

The Bill and Melinda Gates Foundation through the Accelerated Genetic Gain in Rice (AGGRi) Alliance project sponsored and funded this work.

# Additional information

Additional file 1:R scripts for the genomic prediction analysis pipeline currently used at IRRI. Data from the irrigated breeding program are provided as a real case example.

# Figures

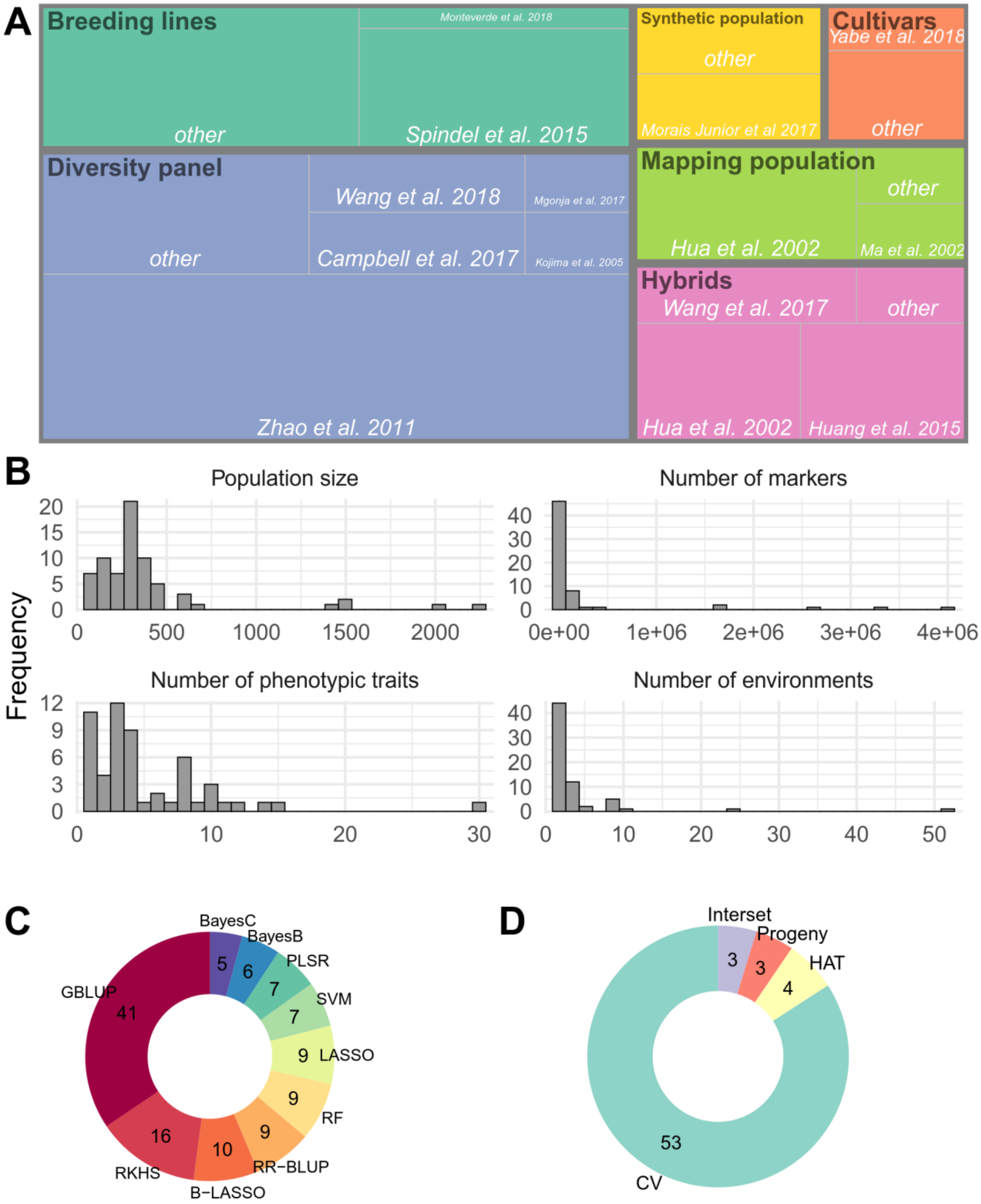
A
Breeding lines
Monteverde et al. 2018
other
Spindel et al. 2015
Synthetic population
other
Morais Junior et al 2017
Cultivars
Yabe et al. 2018
other
Diversity panel
Wang et al. 2018
Mgonja et al. 2017
other
Campbell et al. 2017
Kojima et al. 2005
Zhao et al. 2011
Mapping population
other
Hua et al. 2002
Ma et al. 2002
Hybrids
Wang et al. 2017
other
Hua et al. 2002
Huang et al. 2015
B
Frequency
Population size
Number of markers
Number of phenotypic traits
Number of environments
C
BayesC
BayesB
PLSR
SVM
LASSO
RF
RR−BLUP
B−LASSO
RKHS
GBLUP
41
5
6
7
7
9
9
9
10
16
D
Interset
Progeny
HAT
CV
3
3
4
53

**Figure 1**: Summary of the literature on genomic prediction of rice. It represents the information detailed in Table 1. (A) Treemap of the types of populations used to train genomic prediction models and the associated references for studies which were based on already published datasets. (B) Histograms of the important characteristics of the datasets: the size of the population, the number of phenotypic traits, the number of environments in which the traits were measured (year, season or location) and the number of molecular markers used for genomic predictions. (C) Circle diagram of the ten most used prediction models over the 54 studies. (D) Circle diagram of the validation strategy used to assess the accuracy of prediction models: cross-validation (CV), HAT method, inter-set validation and progeny validation.

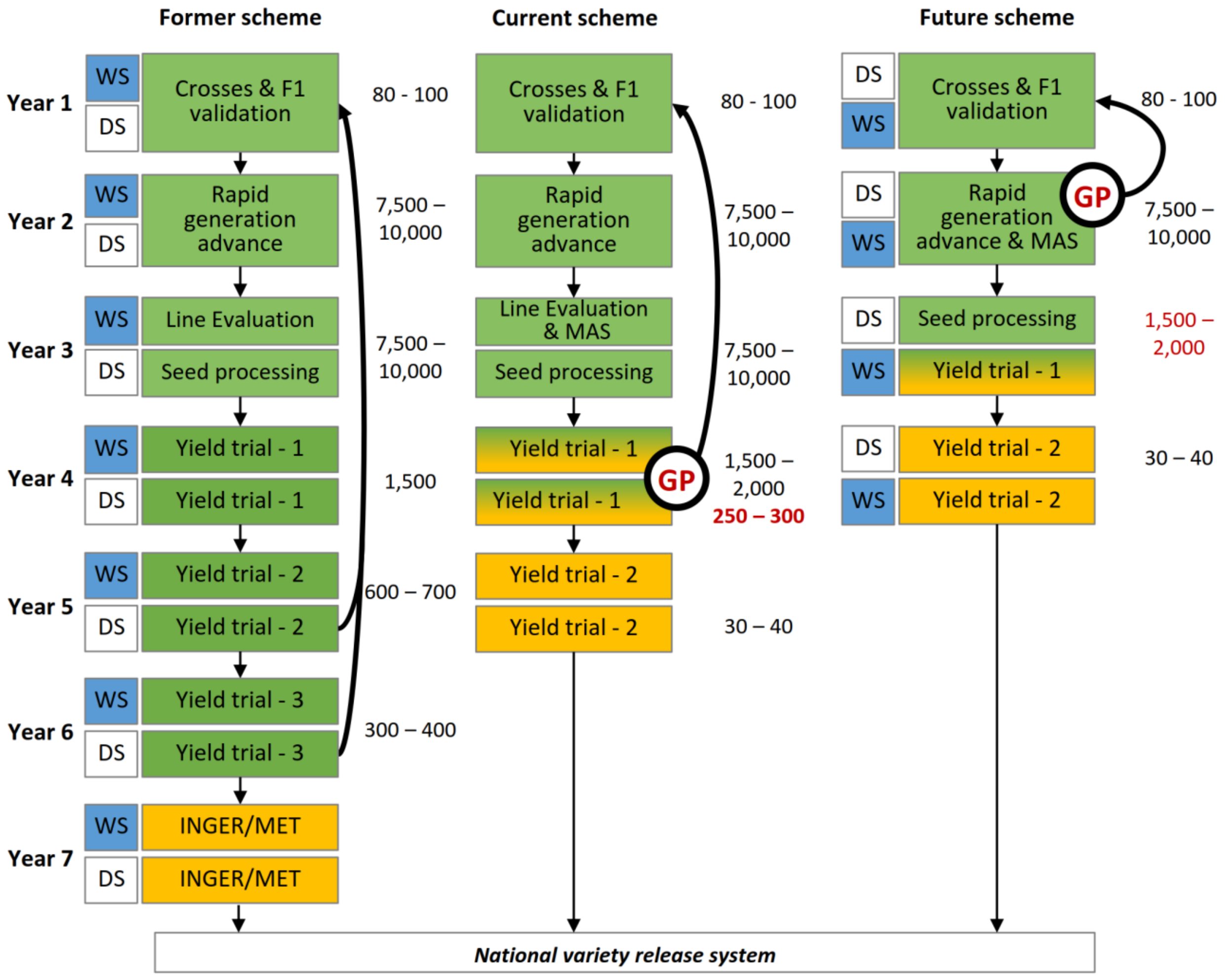

**Figure 2**. Former, current and future breeding schemes at IRRI for irrigated systems. The evolution between the schemes is characterized by the integration of genomic prediction (GP) and a reduction of the breeding cycle length. The genomic prediction is indicated in red with the associated number of individuals being phenotyped in the regions to update the model. The color of the steps corresponds to the location of the activities: green in the Philippines and yellow in the regions with the partners. The years and the seasons (WS: wet season, DS: dry season) are indicated on the left side. The numbers on the right indicate the population size of each step. The black thick arrows indicate the recycling of the best lines as parents. MAS: marker assisted selection using 10-20 trait markers mostly related to disease resistance. INGER: International network for genetic evaluation of rice led by IRRI.

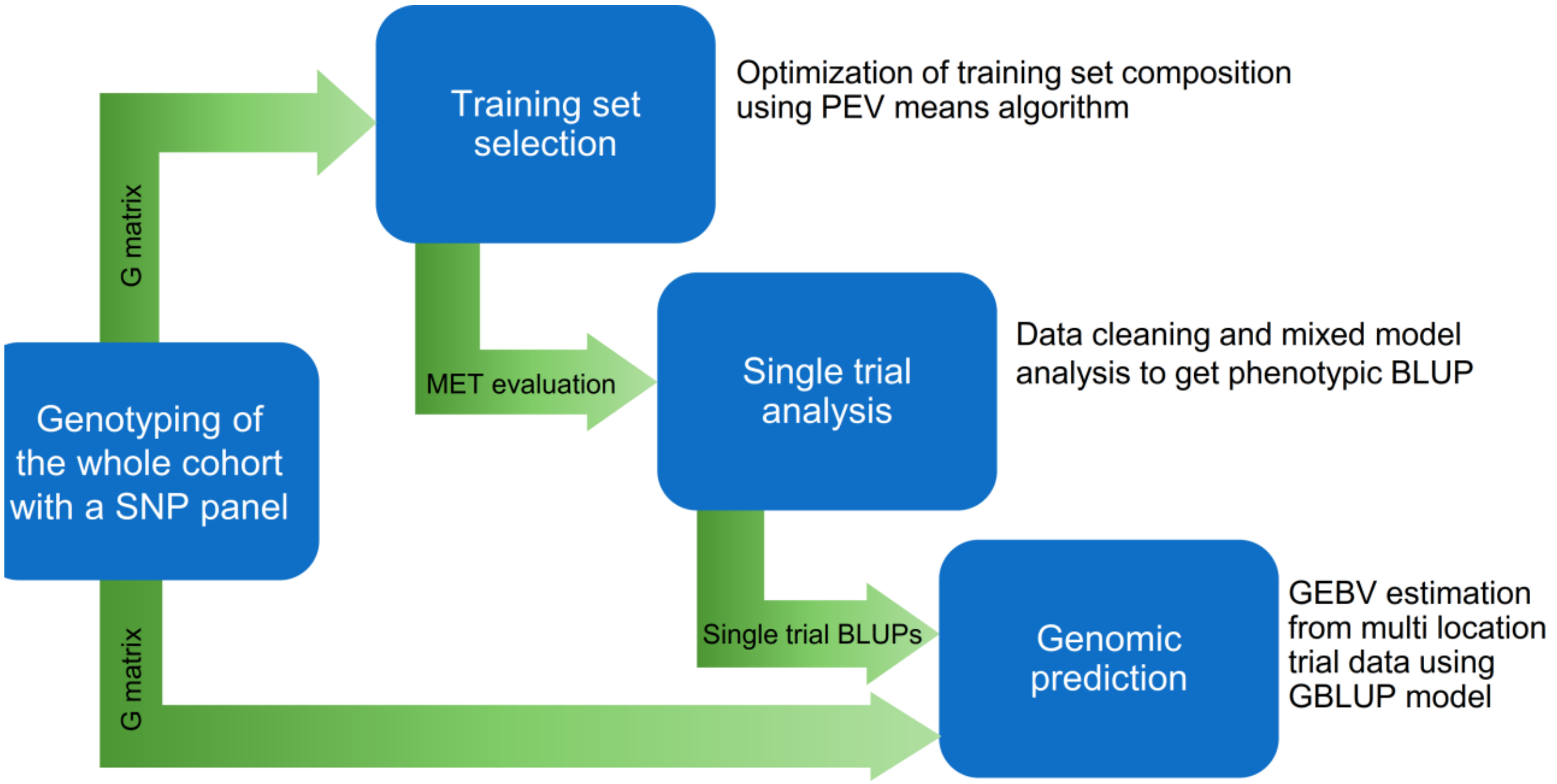

**Figure 3.** The data analysis flowchart represents the routine steps that are performed for every breeding cycle at IRRI's breeding program. The whole cohort (first stage yield trial) is first genotyped with a SNP panel and the data is used to select a training population (subset of the whole cohort). The training population is then evaluated in multi-environment trials (MET). The single trials are analyzed with a mixed model that takes into account the experimental design. The single trial BLUPs combined with the marker information of the whole cohort are then used to compute the genomic estimated breeding values (GEBV) of the lines.

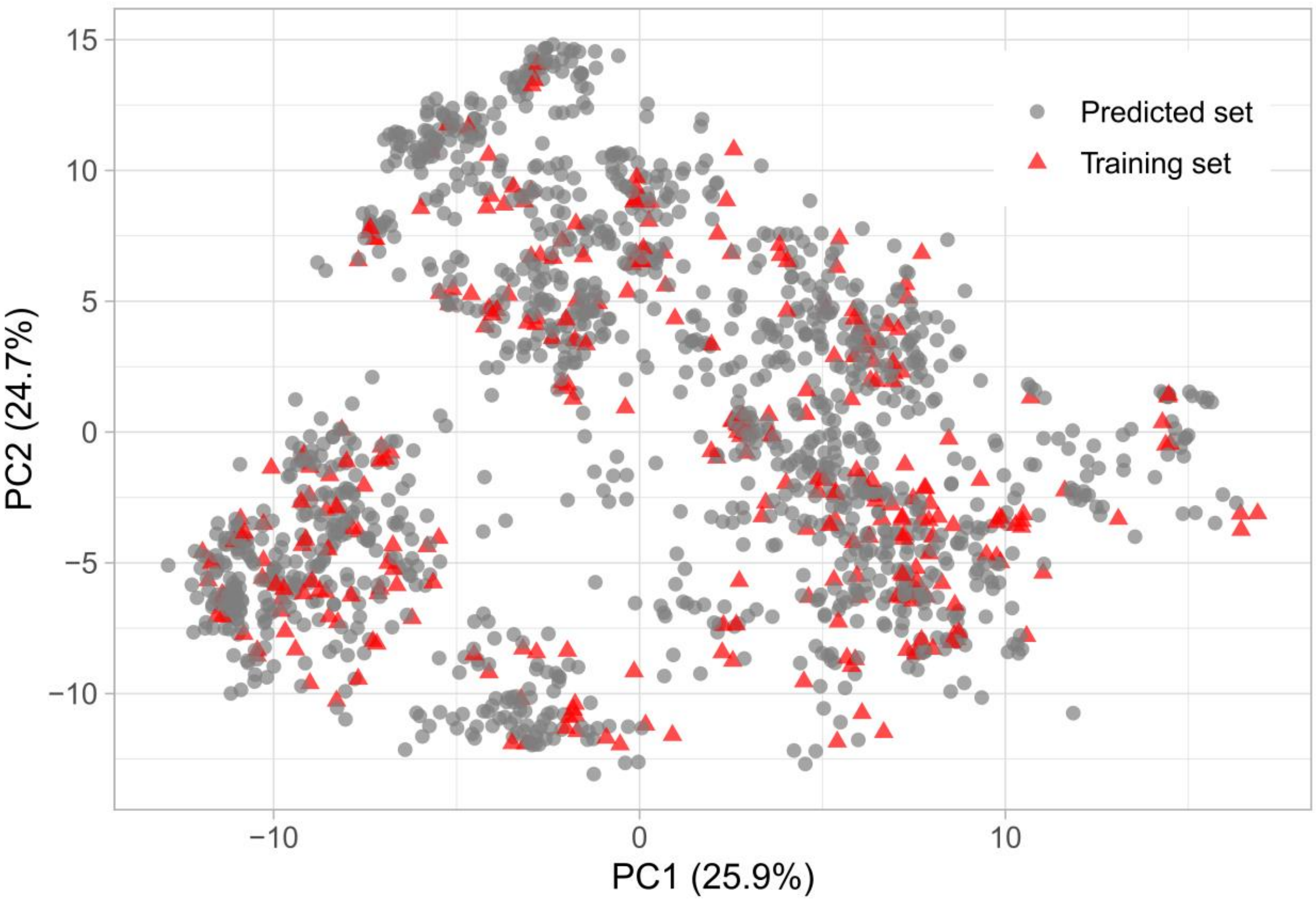

**Figure 4.** Principal component analysis using the molecular marker data on all breeding lines used for genomic prediction. The black triangles represent the lines selected to form the training set using the optimization method of Akdemir et al. [39]. The remaining lines (in red circles) compose the predicted set.

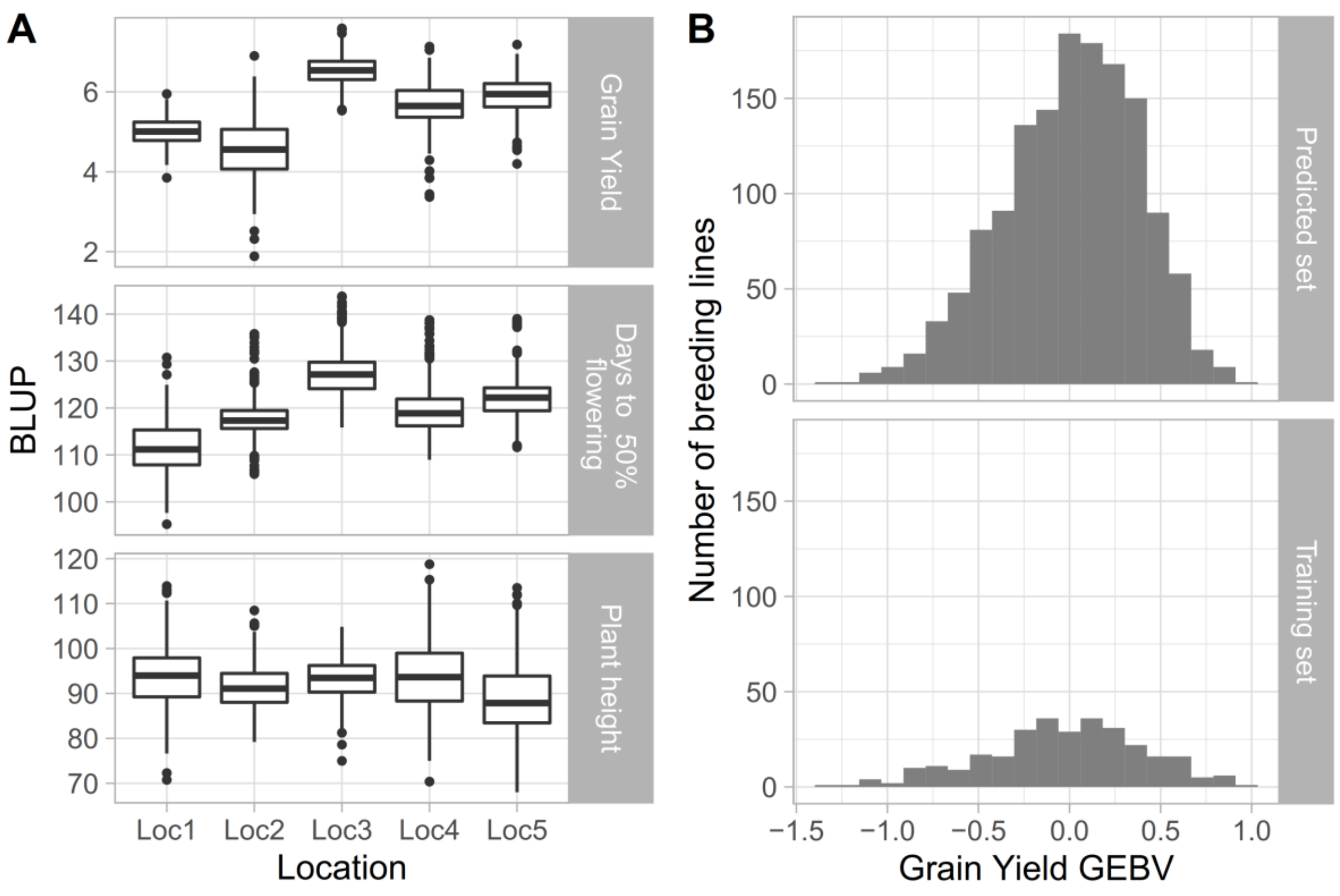

**Figure 5.** Results from the single trial analysis and genomic prediction analysis. Panel A shows the boxplot of BLUP values for grain yield, days to 50% flowering and plant height from the 5 partner trial locations. Panel B presents the distribution of grain yield GEBV of the predicted and training set lines. The results were obtained using *asreml*.

# Tables

**Table 1**: Studies on genomic prediction in rice. When multiple data sets were used in a study, the information is reported only for the rice dataset.

| Reference | Population Type | Population Size | Number of Traits | Number of Markers | Prediction models | Type of validation | Accuracy | Main objective |
|---|---|---|---|---|---|---|---|---|
| Guo et al. [36] | Diversity panel [94] | 413 | 30 | 36901 | GBLUP | Cross-validation (k-fold) | 0.21 - 0.84 | Empirically evaluate the impact of population structure on the accuracy of genomic prediction using cross-validation experiments on the genomic prediction model (GBLUP) |
| Xu et al. [37] | Hybrids [95] | 278 - 105 | 4 | 1619 | GBLUP, LASSO, SSVS | Cross-validation (k-fold) - Interset validation | 0 - 0.69 | Investigate the effect of nonadditive variances on the efficiency of genomic prediction for hybrid performance |
| Zhang et al. [38] | Diversity panel [94] | 413 | 11 | 36901 | GBLUP, BayesB, BLUP\|GA | Cross-validation (k-fold) | 0.51 - 0.85 | Incorporate trait-specific genomic relationship matrices utilizing existing knowledge of genetic architectures in form of significant QTL regions obtained in independent association studies into genomic prediction models to improve the accuracy |
| Akdemir et al. [39] | Diversity panel [94] | 413 | 6 | 36901 | RR-BLUP | Cross-validation (subsampling) | 0.2 - 0.8 | Designof a training population to maximize the accuracy of the genomic prediction models using a genetic algorithm |
| Blondel et al. [40] | Diversity panel [94] | 335 | 14 | 1311 | RF,Ordinal McRank, RankSVM, GBRT, RKHS RR, LambdaMART, B-LASSO EB-LASSO, MIX, SSVS, BayesC, wBSR | Cross-validation (k-fold) | 0.68 - 0.72 | Formulate genomic prediction as the problem of ranking individuals according to their breeding value to employ machine learning methods for ranking |
| Grenier et al. [41] | Synthetic breeding population | 343 | 4 | 8336 | B-LASSO,B-RR, GBLUP, RR-BLUP, LASSO | Cross-validation (k-fold) | 0.12 - 0.54 | Investigate the effect of key factors (training population size and composition, number of markers, model) on the accuracy of genomic prediction |
| Isidro et al. [42] | Diversity panel [94] | 413 | 4 | 36901 | RR-BLUP | Cross-validation (subsampling) | 0.22 - 0.73 | Compare the performance of different optimization criteria in the presence of population structure and evaluate how population structure interacts with these criteria in the choice of the training population |
| Iwata et al. [43] | Two diversity panels | 179 - 386 | 1 | 3,254 - 36,901 | GBLUP, RKHS, PLSR, KPLSR | Cross-validation (leave one out, k-fold) | 0.4 - 0.64 | Propose a method for predicting rice grain shape delineated by elliptic Fourier descriptors based on genome-wide marker polymorphisms |
| Onogi et al. [44] | Cultivars | 110 - | 8 | 3102 | GBLUP, RKHS, LASSO, Elastic net, RF, B-LASSO, EB-LASSO, BSR | Cross-validation (k-fold) | 0.40 - 0.84* | Assess the performance of eight different methods on the accuracy of prediction using real and simulated data |
| Spindel et al. [45] | Breeding lines | 332 | 3 | 73147 | RR-BLUP, B-LASSO, RKHS, RF, MLR, PBLUP | Cross-validation (subsampling) | 0 - 0.63* | Investigate the effect of the number of markers, the model and the trait architecture on the accuracy of genomic prediction |
| Bustos-Korts et al. [46] | Diversity panel [94] | 413 | 3 | 26259 | GBLUP, QGBLUP, RKHS | Cross-validation (subsampling) | 0.28 - 0.81 | Evaluate different methods to optimize the training population and their interaction with prediction models |
| Jacquin et al. [47] | Breeding lines and Diversity panel | 230 - 167 - 188 | 15 | 22,691 - 16,444 - 38,390 | LASSO, GBLUP, SVM, RKHS | Cross-validation (k-fold) | 0.12 - 0.70 | Provide a clear and unified understanding of parametric statistical and kernel methods, used for genomic prediction, and to compare some of these in the context of rice breeding |
| Onogi et al. [48] | Mapping population (Ma et al. 2002) | 174 | 1 | 162 | EB-LASSO, EB-LASSO + crop model | Cross-validation (leave one out) | 0.87 - 0.97 | Predict heading date by coupling genomic prediction and crop model |
| Spindel et al. [49] | Breeding lines | 332 | 3 | 58318 | GBLUP, RR-BLUP, B-LASSO, RKHS, RF, MLR | Cross-validation (subsampling) | 0 - 0.65* | Assess the potential of introducing fixed variables identified using de novo GWAS into GS models to improve prediction accuracy, and also consider the contribution of multi-location field trials to GS prediction accuracy |
| Campbell et al. [50] | Diversity panel [94] | 360 | 1 | 36901 | GBLUP | Cross-validation (k-fold) | 0.39 - 0.73 | Asses the accuracy of genomic prediction of shoot growth dynamic |
| Gao et al. [51] | Breeding lines [45] | 315 | 3 | 58227 | GBLUP (10 relationship matrices) | Cross-validation (k-fold) | 0.24 - 0.57 | Incorporate gene annotation into relationship matrices to improve accuracy of genomic prediction |
| Matias et al. [52] | Breeding lines [45] | 270 | 2 | 39915 | B-RR, BayesB, B-LASSO | Cross-validation (subsampling) | 0.26 - 0.42 | Use of haplotype blocks as multiallelic markers to improve the accuracy of genomic prediction |
| Morais et al. [53] | Synthetic breeding population | 174 | 8 | 6174 | GBLUP (5 relationship matrices) | Cross-validation (subsampling) | 0.31 - 0.68 | Assess the relevance of additive and nonadditive genetic effects on the predictive accuracy |
| Wang et al. [54] | Hybrids | 575 | 8 | 3299150 | GBLUP (univariate and multivariate) | Cross-validation (k-fold) | 0.40 – 0.86 | Investigate the performance of multivariate models including dominance for predicting phenotypes of rice hybrids benefiting from joint analysis with auxiliary traits or with the phenotypes observed in other environments |
| Xu et al. [55] | Hybrids | 1495 | 10 | 1654030 | GBLUP | HAT, Cross-validation (k-fold) | 0.40 - 0.88 | Develop an alternative method, the HAT method, to replace cross-validation in the context of genomic prediction |
| Ben Hassen et al. [56] | Diversity panel Breeding lines | 284 - 97 | 3 | 43686 | GBLUP, RKHS (univariate and multivariate) | Cross-validation (subsampling) - Progeny validation | -0.12 - 0.96 | Explore the feasibility of genomic selection for the adaptation of rice to alternate wetting and drying in the framework of a pedigree breeding scheme |
| Ben Hassen et al. [57] | Diversity panel Breeding lines | 284 - 97 | 3 | 43686 | GBLUP, RKHS, BayesB | Cross-validation (subsampling) - Progeny validation | 0.23 - 0.65 | Investigate the impact of the size and the composition of the training population that maximize the accuracy of prediction of phenotype of progeny lines |
| Campbell et al. [58] | Diversity panel [50] | 357 | 1 | 33674 | GBLUP | Cross-validation (k-fold, subsampling) | 0.4 - 0.89 | Examine the advantage of utilizing random regression models for longitudinal phenotypes over single end-point measurement in the context of genomic prediction |
| Du et al. [59] | Mapping population [95] | 210 | 4 - 1,000 - 24,973 | 1619 | RR-BLUP, PCR, PLSR | Cross-validation (k-fold) - HAT | 0.12 - 0.76 | Evaluate the advantages of principal components regression over partial least square regression for genomic prediction of agronomic, metabolomic and transcriptomic traits |
| Gao et al. [60] | Breeding lines [45] | 315 | 3 | 58227 | GBLUP (7 reationship matrices) | Cross-validation (k-fold) | 0.24 - 0.56 | Incorporate gene annotation information into genomic prediction models by constructing haplotypes with SNPs mapped to genic regions to improve accuracy |
| Mathew et al. [61] | Diversity panel [94] | 371 | 1 | 36901 | GBLUP (multivariate) | Cross-validation (k-fold) | 0.49 - 0.77 | Study the impact of different residual covariance structures on genomic prediction ability using different models to analyze multi-environment trial data |
| Monteverde et al. [62] | Breeding lines | 309 - 327 | 5 | 44,598 - 92,430 | GBLUP, RHKS (multivariate) | Cross-validation (subsampling) | 0.30 - 0.88 | Compare the effect on prediction accuracy of different multi-environment models and different training populations |
| Morais Júnior et al. [63] | Synthetic breeding population [53] | 174 | 8 | 6174 | ABLUP, GBLUP, AGBLUP HBLUP, BayesC, B-LASSO, PLSR, RF, RKHS | Cross-validation (subsampling) | 0.23 - 0.76 | Compare prediction models to identify the most accurate and develop low-risk genomic selection methods for use in rice breeding |
| Morais Júnior et al. [64] | Synthetic breeding population [53] | 667 - 174 | 3 | 6174 | Bayesian HBLUP (multivariate with environmental covariates) | Cross-validation (subsampling) | -0.15 - 0.9 | Evaluate single step models incorporating environmental covariates and the importance of main effects and interaction components for the prediction of phenotypic responses |
| Xu et al. [65] | Hybrids [54] | 575 | 8 | 2561889 | GBLUP, PLSR, LASSO, BayesB, SVM, RKHS | Cross-validation (k-fold) | 0.15 - 0.88 | Evaluate effects of statistical methods, heritability, marker density and training population size on prediction for hybrid performance |
| Yabe et al. [66] | Cultivars | 123 | 1 | 42508 | GBLUP, PLSR | HAT, Cross-validation (leave one out) | 0.22 - 0.53 | Develop a method to describe grain weight distribution and evaluate the efficiency of genomic prediction for the genotype-specific parameters of grain weight distribution |
| Arbelaez et al. [67] | Breeding lines | 353 | 3 | 965 | ABLUP, RR-BLUP, BayesA, BayesB, BayesC, B-LASSO, RKHS | Cross-validation (k-fold) | 0.36 - 0.71 | Assess the effectiveness of a genotyping platform of a thousand highly informative SNP sites for genomic prediction in *indica* based breeding programs |
| Azodi et al. [68] | Breeding lines | 327 | 3 | 73147 | RR-BLUP, B-RR, BayesA, BayesB, B-LASSO, SVM, RF, GTB, ANN, CNN | Cross-validation (k-fold) | 0.25 - 0.65 | Compare the performance of different prediction models including artificial neural networks using available datasets |
| Berro et al. [69] | Breeding lines [62] | 317 - 327 | 1 | 44,598 92,430 | GBLUP | Cross-validation (k-fold) | 0.37 - 0.80 | Compare strategies for optimizing the training set for genomic prediction models |
| Bhandari et al. [70] | Diversity panel | 280 | 3 | 215242 | GBLUP, RKHS | Cross-validation (subsampling) | 0.23 - 0.81 | Investigate on the effectiveness of trait-specific marker selection and of multi-environment prediction models in improving the accuracy of genomic predictions for drought tolerance in rice |
| E Sousa et al. [71] | Breeding lines [45] | 270 | 2 | 39811 | GBLUP, RKHS | Cross-validation (subsampling) | 0.18 - 0.31 | Compare the effect of two strategies to obtain markers subsets and their effect on prediction accuracy, bias and the relative efficiency of a main genotypic effect model |
| Frouin et al. [72] | Diversity panel | 228 - 95 | 2 | 22370 | GBLUP, BayesA, RKHS | Cross-validation (subsampling) - | 0.23 - 0.54 | Explores the feasibility of genomic selection to improve the ability of rice to prevent arsenic uptake and accumulation in the edible grains |
| | Breeding lines | | | | | Interset validation | | |
| Guo et al. [73] | Hybrids | 1439 | 4 | 1654030 | GBLUP | Cross-validation (subsampling) | 0.59 - 0.77* | Optimize the training population for the genomic prediction of hybrid performance using design-thinking and data-mining techniques |
| Hu et al. [74] | Mapping population [95] | 210 | 4 - 1,000 - 24,973 | 1619 | multilayered-LASSO | Cross-validation (k-fold) | 0.16 - 0.76 | Evaluate a novel strategy of genomic prediction called multilayered least absolute shrinkage and selection operator (ML-LASSO) by integrating multiple omic data into a single model that iteratively learns three layers of genetic features supervised by observed transcriptome and metabolome |
| Huang et al. [75] | Diversity panel [94] | 161 - 162 | 1 | 66,109 29,030 | RR-BLUP, GBLUP (multivariate), BayesA, BayesC | Cross-validation (k-fold) | 0.15 - 0.80 | Assess the utility of genomic prediction in improving rice blast resistance |
| Lima et al. [76] | Diversity panel [94] | 370 | 7 | 36901 | GBLUP, Delta-p | Cross-validation (k-fold) | 0.27 - 0.83 | Propose the Delta-p (method based on the genetic distance between two subpopulations, using the concepts of changes in allele frequency due to selection and the genetic gain theory) and Delta-p/G-BLUP index, and to compare it with the traditional G-BLUP method |
| Monteverde et al. [77] | Breeding lines | 309 - 327 | 4 | 44,598 92,430 | GBLUP, PLSR (multivariate environmental covariates) | Cross-validation (subsampling) | 0.10 - 0.90* | Use molecular marker data and environmental covariates simultaneously to predict rice yield and milling quality traits in untested environments |
| Ou et al. [78] | Diversity panel [94] | 404 | 10 | 30315 | RR-BLUP | Cross-validation (subsampling) | 0.09 - 0.78* | Propose a new criterion derived from Pearson's correlation between GEBVs and phenotypic values of a test set to determine a training set for genomic prediction |
| Suela et al. [79] | Diversity panel [94] | 352 | 9 | 36901 | Delta-p, GBLUP, BayesC, B-LASSO | Cross-validation (k-fold) | 0.10 - 0.83 | Evaluate the Delta-p/BLASSO and Delta-p/BayesCpi genomic indexes and compare them to the Delta-p/G-BLUP index in terms of prediction efficiency of additive genomic values |
| Wang et al. [80] | Hybrids, Mapping population [95] | 210 - 278 | 4 - 1,000 - 24,973 | 1619 | LASSO, GBLUP, SVM, PLSR | Cross-validation (k-fold) | 0.1 - 0.70* | Prove the concept that trait predictability may be optimized by using superior prediction models and selective omic datasets |
| Wang et al. [81] | Hybrids [54] | 575 | 8 | 61836 | GBLUP | Cross-validation (k-fold) | 0.07 - 0.15 | Combine selection index with genomic prediction method to predict hybrid rice for a more accurate and comprehensive selection |
| Baba et al. [82] | Diversity panel [50, 95] | 357 | 2 | 34993 | Random regression (uni and multivariate) | Cross-validation (subsampling) | 0.17 - 0.91* | Demonstrate the utility of a multi-trait random regression models for genomic prediction of daily water usage in rice through joint modeling with shoot biomass |
| Banerjee et al. [83] | Breeding lines [45] | 315 | 3 | 73147 | RR, LASSO, SVM, Bagging, RF, AdaBoost, XGBoost | Cross-validation (k-fold) | 0.10 - 0.67 | Compare linear and non-linear prediction methods and assess the efficiency of dimensionality reduction approaches using rice as an example |
| Cui et al. [84] | Hybrids | 1,495 - 100 | 106 | 102795 | GBLUP (multivariate) | Cross-validation (k-fold) - Interset | 0.35 - 0.92 | Use genomic best linear unbiased prediction to predict hybrid performances using cross-validation and inter-set validation |
| Grinberg et al. [85] | Diversity panel [30] | 2265 | 12 | 101595 | LASSO, RR, GBLUP, GBM, RF, SVM | Cross-validation (k-fold) | 0.14 - 0.70 | Compare standard machine learning methods with the state of-the-art classical statistical genetics method: GBLUP |
| Jarquin et al. [86] | Cultivars [66] | 112 | 1 | 408372 | GBLUP | Cross-validation (subsampling) | 0.41 - 0.93 | Propose two novel methods for predicting days to heading in rice of tested and untested genotypes in unobserved environments in a precise and accurate way |
| Schrauf et al. [87] | Diversity panel [30] | 2018 | 1 | 4000000 | GBLUP 3 relationship matrices) | Cross-validation (k-fold) | 0.16 - 0.83* | Explore how the difference in predictive ability of epistatic models and additive models is related to the density of the markers used for predictions, and put observations in the context of phantom epistasis |
| Toda et al. [88] | Mapping population | 123 | 1 | 315 | GBLUP, LASSO,RR, RKHS, RF (integration with crop model) | Cross-validation (k-fold, subsampling) | 0.40 - 0.68* | Develop models to predict the biomass of rice with the integration of observed phenotypic data of growth-related traits, whole-genome marker genotype, and environmental data. |
| Xu et al. [89] | Hybrids, Mapping population [95] | 210 - 278 | 4 - 1,000 - 24,973 | 1619 | GBLUP (different relationship matrices) | HAT - Progeny validation | 0.20 – 0.80* | Integrate parental phenotypic information into various multi-omic prediction models applied in hybrid breeding of rice and compared the predictabilities of 15 combinations from four sets of predictors from the parents, that is genome, transcriptome, metabolome and phenome |

* read from graphics not mentioned in the text